\documentclass[journal,onecolumn]{IEEEtran}

\ifCLASSINFOpdf
\else
   \usepackage[dvips]{graphicx}
\fi

\usepackage{amsmath,amsfonts}
\usepackage{algorithmic}
\usepackage{algorithm}
\usepackage{array}
\usepackage[caption=false,font=normalsize,labelfont=sf,textfont=sf]{subfig}
\usepackage{textcomp}
\usepackage{stfloats}
\usepackage{url}
\usepackage{verbatim}
\usepackage{graphicx}
\usepackage{cite}
\usepackage{hyperref}
\usepackage{orcidlink}
\ifpdf
\hypersetup{
  colorlinks=true,
  linkcolor=blue,
  citecolor=blue,
  urlcolor=blue,
}
\fi

\newcommand{\eg}{{\em e.g., }}
\newcommand{\ie}{{\em i.e., }}
\newcommand{\etal}{{\em et al.,~}}

\newcommand{\pkg}[1]{\texttt{#1}}

\begin{document}
\title{A faster algorithm for efficient longest common substring calculation for non-parametric entropy estimation in sequential data}

\author{Bridget Smart \orcidlink{0000-0002-0910-9470}, Max Ward \orcidlink{0000-0001-9114-7339}, and Matthew Roughan\orcidlink{0000-0002-7882-7329},
\thanks{Bridget Smart is with Mathematical Institute, University of Oxford, United Kingdom. Max Ward is with The University of Western Australia, Australia. Matthew Roughan is with The University of Adelaide, Australia.}
\thanks{
The authors would like to thank Lewis Mitchell for his contributions and support in conceptualising this work.}
\thanks{B. Smart w a Westpac Future Leaders Scholarship. M.~Roughan is supported by the Australian Government through the Australian Research Council’s Discovery Projects funding scheme (project DP210103700). There was no additional external funding received for this study.}}



\maketitle
\begin{abstract} 

Non-parametric entropy estimation on sequential data is a fundamental tool in signal processing, capturing information flow within or between processes to measure predictability, redundancy, or similarity. Methods based on longest common substrings (LCS) provide a non-parametric estimate of typical set size but are often inefficient, limiting use on real-world data. We introduce LCSFinder, a new algorithm that improves the worst-case performance of LCS calculations from cubic to log-linear time. Although built on standard algorithmic constructs---including sorted suffix arrays and persistent binary search trees---the details require care to provide the matches required for entropy estimation on dynamically growing sequences. We demonstrate that LCSFinder achieves dramatic speedups over existing implementations on real and simulated data, enabling entropy estimation at scales previously infeasible in practical signal processing.

\end{abstract}


\section{Introduction}

\IEEEPARstart{C}{alculating} the longest common substring is an established problem in computer science and a crucial step in estimating the Shannon entropy rate of a text and cross entropy rate between a pair of texts \cite{wyner_asymptotic_1989,bergroth_survey_2000, paterson_longest_1994}. The entropy estimator proposed by Kontoyiannis~\etal \cite{kontoyiannis_nonparametric_1998}, for instance, uses the ratio between the database length and the length of the shortest unseen substring to estimate the self entropy rate of a sequence. The Kontoyiannis~\etal estimator is one example of the Lempel-Ziv family of entropy estimators \cite{zivUniversalAlgorithmSequential1977,lempelComplexityFiniteSequences1976}. This family of estimators uses the longest common substring and has been widely applied, for instance, to construct information-theoretic models for online social networks \cite{wyner_asymptotic_1989,southInformationFlowEstimation2022,bagrow_information_2019,pond_complex_2020,he_identifying_2013}. Existing implementations scale poorly with sequence length and number of sources when applied to sequential non-parametric entropy estimation, especially when applied to long sequences or many processes as these existing approaches either calculate match lengths across all indices in the two strings or apply algorithms designed for data compression rather than entropy estimation~\cite{southInformationFlowEstimation2022,benderNewAsymptoticBounds1991,en-huiyangSimpleUniversalLossy1996}. 



The main contribution of this work is an improved algorithm, LCSFinder, for efficiently calculating the length of the longest common substring between strings $S$ of length $N_S$ and $T$ of length $N_T$ under the condition that the lengths may be dynamically growing. Achieving a theoretical worst-case computational complexity of $O(N \log^2 N)$ when $N_S = N_T = N$, our method greatly improves upon existing approaches, which have worst-case complexity of $O\big(N^3\big)$. To achieve this, LCSFinder leverages sorted suffix arrays and persistent binary search trees to efficiently determine substring match lengths.



We also experimentally verify the improvement in {\em expected} complexity of LCSFinder using both simulated and real world datasets representative of a typical use cases. In computational social science, large datasets frequently contain millions of events from thousands to millions of unique sources. To evaluate how the LCSFinder algorithm will perform for this application, we use an X (formerly Twitter) dataset containing 2,846,284 tweets from 123 accounts from South \textit{et al.}~\cite{southInformationFlowEstimation2022}. We compare LCSFinder against an existing method for measuring information flow between account pairs, implemented in \pkg{ProcessEntropy} \cite{southInformationFlowEstimation2022}. 


LCSFinder has been released as a standalone package\footnote{Available at: \url{https://pypi.org/project/LCSFinder}}, with complete code available at \url{https://github.com/bridget-smart/LCSFinder}. In addition, a heuristic algorithm that achieves excellent practical performance at the cost of slow worst-case performance is included. The updated algorithm has also been integrated into an existing entropy package which uses the non-parametric entropy estimator, with code available in the package \pkg{ProcessEntropy}\footnote{Available at \url{https://pypi.org/project/ProcessEntropy}.}. 

\section{Background}

\subsection{Notation}

Let $S$ and $T$ denote the source and target sequences of lengths $N_S$ and $N_T$, respectively. We write $S_a^b = (S_a, S_{a+1}, \ldots, S_b)$ for a substring of $S$, and similarly for $T$. Each symbol is drawn from a finite alphabet $\mathcal{V}$ of size $|\mathcal{V}|$. When applied to language, $\mathcal{V}$ represents the vocabulary, with each value representing a distinct word and $|\mathcal{V}|$ representing the true vocabulary size.

For cross-entropy estimation, we compare the \textit{history} of $S$ to the \textit{future} of $T$ at a given instant $t$. Specifically, for indices $i$ and $j$ corresponding to positions in $S$ and $T$, we define $\Lambda_i^j$ as the length of the shortest substring in $T$ beginning at $i$ that does not appear in the prefix of $S$ up to position $j$,  
\[
\Lambda_i^j = \max_{\ell}\{\,S_m^{m+\ell} = T_j^{j+\ell}\;|\; m+\ell < i,\; j+\ell \le N_T \,\} + 1.
\]
This quantity provides the match length used by the non-parametric entropy estimator. When applied over time or across many sequences, these match lengths form the basis for computing self-entropy or cross-entropy rates.

\subsection{Entropy Calculation}

The {\em entropy rate} of a string or sequence is a quantity that describes the long-term average of new information per symbol in the string and is directly related to the information content of the string and represents an asymptotic lower bound on the lossless compression ratio. Entropy rate is of great interest across a variety of fields from signal processing, cryptography, compression, computational neuroscience and applied mathematics. 



While frequently used, the Shannon entropy rate does not account for dependencies that exist within most sequences of interest, \eg text. There are many alternative approaches motivated by the Asymptotic Equipartition Theorem (AEP)~\cite{cover2006} which are robust and able to measure dependencies in sequential data. 

An example of this is the estimator proposed by Kontoyiannis~\etal\cite{kontoyiannis_nonparametric_1998},
$$
\hat{K}(n) = \dfrac{1}{n} \sum_{i=1}^n \dfrac{\Lambda_i^i}{\log n} 
\stackrel{n \rightarrow \infty}{\longrightarrow} \dfrac{1}{H},
$$
where $\Lambda_i^i$ denotes the length of the shortest substring starting from position $i$ which does not appear in the previous $i$ symbols and $n$ is the length of the sequence\footnote{In fact, Kontoyiannis~\etal propose three similar estimators based on Cesaro averages of match lengths. We use the one shown here to illustrate the manner in which matches are used.}. This estimator converges almost surely \cite{ornstein_entropy_1993, wyner_asymptotic_1989, kontoyiannis_nonparametric_1998}. For a sequence $X$ we define 
\begin{align*}
\Lambda_i^i &= \max \left\{ \ell \mid 0 \leq \ell \leq n, X_1^{\ell-1} = X_{-j}^{-j+\ell-1},  \ell \leq j \leq n \right\} + 1\\
&=\min \left\{ \ell^* \mid \ell^*>0, X_1^{{\ell^*}-1} \neq X_{-m}^{-m+{\ell^*}-1},  1 \leq m \leq n \right\}.
\end{align*}

The cross entropy between two sources describes the amount of information required to describe one source given the other. For instance, the cross entropy between probability distributions $p$ and $q$, $H(p~||~q)$, is defined as,
$$
H(p~||~q) = - \sum_x p(x) \log q(x).
$$
This measure is not symmetric, as it assumes complete knowledge of $q$ and only partial knowledge of $p$. In many applications symmetry would be counter-indicated because we aim to measure the influence (the past of) a source has on (the future of) the target, so the cross-entropy measure is ideal.

The estimator proposed by Kontoyiannis~\etal was extended in work by Bagrow \textit{et al.}~\cite{bagrow_information_2019} to calculate cross entropy. This cross entropy estimator uses pattern match lengths between the history of one sequence and the future of the other to gain insight into the quantity of information shared between the two processes over time. 

When this non-parametric cross entropy estimator is used to model information flows over online social networks, the pairwise cross entropy represents the directed influence between two users \cite{bagrow_information_2019,wuInformationFlowSocial2004,vosoughiSpreadTrueFalse2018}. This body of work looks to understand the flow of information across social ties as a measure for how predictable a user's behaviour is given their connections. In applications with many distinct generating processes, this estimator involves calculating the longest common match length pairwise between each process or user, so the scalability and complexity of this algorithm plays an important role in the feasibility of this technique for large datasets. When entropy is calculated in real time, these sequences can increase in length, so estimation tools which can be calculated quickly and efficiently are ideal.

\subsection{Existing work}

Apart from the work above, existing work on entropy estimation in the computational social sciences predominately uses parametric estimators based on stochastic models, \eg Markov models \cite{cover2006}, or sequential non-parametric entropy estimators often motivated by the compression bound of entropy measures \cite{lempelComplexityFiniteSequences1976,zivUniversalAlgorithmSequential1977,en-huiyangSimpleUniversalLossy1996,benderNewAsymptoticBounds1991,williamsExtremelyFastZivLempel1991,beluRoLZReducedOffset2019}. These are either (i) quite inaccurate when sequences contain complex structures that are hard to model by a finite state machine such as in natural language, and/or (ii) do not provide a mechanism to generalise to cross entropy measurement. The Kontoyiannis~\etal non-parameteric entropy estimator \cite{kontoyiannis_nonparametric_1998}, and other non-parametric estimators which consider match lengths, such as those derived from the work of \cite{wyner_asymptotic_1989,ornstein_entropy_1993}, avoid both issues. However, the requirement to calculate match lengths on long sequences for many pairs of sequences has made these estimators hard to use in practice due to the computational complexity of existing string matching approaches.  




The package \textsc{ProcessEntropy}\footnote{Available at \url{https://pypi.org/project/ProcessEntropy}.}\cite{southInformationFlowEstimation2022} is an open-source implementation of the Kontoyiannis~\etal estimator. It has been updated with the implementation described here, but previously used a brute-force search at each index to calculate the longest common substring for a pair of processes, resulting in worst-case complexity to calculate the longest common substring between a pair of processes, resulting in a worst-case complexity of $O\big(N^3\big)$ for sequences of lengths $N_S=N_T=N$. For long sequences or large numbers of processes, the computational complexity was a major limiting factor. The previous version of the estimator was used on several large online social network datasets  \cite{southInformationFlowEstimation2022,smart2022istandwithputin}, consuming many days of computer time and limiting the scope of the analysis. 


\section[Calculating the LCS]{Calculating the longest common substring}

Given two strings of interest, a source $S$ and a target $T$, the goal of the LCSFinder algorithm is to compute the longest match length between part of the source string, and part of the target string. In many cases, these substrings are selected to represent a history and a future relative to a specific point in time, assuming that symbols in both sequences occur in chronological order. 

Where the symbols in each string are aligned temporally, we may be interested in calculating the entropy between the history of the source process and the future of the target at a common index $i$. In this case, we choose the substrings $S_1^{i-1}$, the history of the source process, and $T_i^{N_T}$, the future of the target process, where $N_T$ is the length of the target sequence, and $i$ represents an instant in time. This represents the case where the index temporally matches between both strings, and a slice at $i$ results in comparing the history of the source string with the future of the target string. The most common case for such would be when we are calculating the entropy rate of a single string, \ie $S = T$.

\begin{figure}
\centering
    \includegraphics[width=0.8\columnwidth]{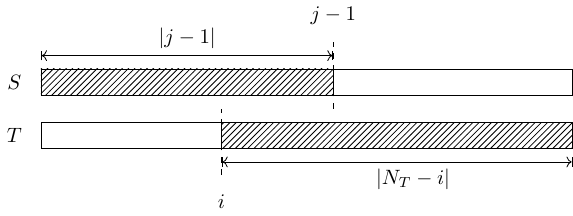}
    \caption{Indices on the strings $S$ and $T$.}
    \label{fig:eg}
\end{figure}

However, we allow that when $S \neq T$ the indices may not be temporally aligned between the source and target string, so we use the generalisation that we are looking for the longest match starting at index $i$ in $T$ which starts and ends prior to index $j$ in $S$ (\autoref{fig:eg}).  We do not put any constraints on the relationship between $i$ and $j$.

When the symbols take non-integer values, the vocabulary of these strings, $S$ and $T$, is mapped to the alphabet $(1,...,V)$, where $|V|$ is the combined vocabulary size of $S$ and $T$.

When used to calculate cross entropy, the estimate at each point in time is averaged to improve the stability of the estimate \cite{kontoyiannis_nonparametric_1998}. For this application, the existing approach performs a match length search for each possible pair of indices. This approach gives a complexity of $O\big(j (j+N_T-i) (N_T-i)\big)$ for a given $i,j$ pair. In the case when $i=j$, this time-averaged estimate has cubic performance for entropy estimation as we calculate the LCS for each value of $i$. 

The primary contribution of this work is the LCSFinder algorithm, a new approach to calculate the longest match lengths between pairs of sequences with significantly improved complexity. By representing the matches as prefixes, the proposed algorithm uses a sorted suffix array \cite{manber1993suffix} to define an ordering on the suffixes and answer longest common prefix queries. Then a persistent binary search tree  \cite{sarnak1986planar} is used to keep track of prefixes of the sorted suffix array order.

Define a \textit{strict query} as those we want to compute for entropy calculation and a \textit{relaxed query} $f(i, j, k)$ as a binary function that decides if there is a match of length $k$ starting at $i$ in $T$ and before $j$ in $S$. Relaxed queries do not force the entire match with the source history to stay inside $S_{1}^{j}$. If we have an efficient algorithm for solving relaxed queries, then we can show there is an efficient algorithm for strict queries. First, we will explain why this is the case. Second, we will show how relaxed queries can be computed efficiently.

Formally, a strict query $g(i,j)$ is the longest match starting at $i$ in $T$ and before $j$ in $S$ such that the match is fully contained inside $T$ and the $S$-prefix $S_{1}^{j-1}$. Observe that if $f(i, j-k, k)$ is true, then $g(i, j)<k$. That is, if we can find a match of length $k$ for the $S$-prefix ending at $j-k$, then the solution to the strict query ending at $j$ must be at least as long as the match of length $k$ we found, since it contains that match.

Consider fixing $i$ and $j$. Next, consider $k \in [0, \min(N_S, N_T))$. The relaxed query function $f(i,j-k,k)$ is a binary step function. That is, if there is a match of length $k$, then there is a match of length $k-1$. Therefore a binary search on $k$ can find the largest $k$ for which there is a match. This answers strict queries in $O(R \log \min(N_S, N_T))$ where $R$ is the complexity of solving a relaxed query. 

Now consider the problem of computing relaxed queries. A sorted suffix array is constructed as precomputation. Define $\Psi = S \cdot \sigma \cdot T$ as the concatenation of $S$, a special symbol $\sigma$, and $T$. The symbol $\sigma$ is used to distinguish between suffixes in $S$ (containing $\sigma$) and those in $T$ (not containing $\sigma$). A sorted suffix array $\mathcal{A}$ is constructed for $\Psi$. The longest common prefix table is constructed alongside $\mathcal{A}$ \cite{manber1993suffix}. Call $\Psi(S, x) \in [1, N_S + N_T + 1)$ the sorted suffix array index the suffix $S_{x}^{N_S}$. Similarly, define $\Psi(T, y)$ for suffixes of $T$. Define $\operatorname{lcp}(a, b)$ as the longest common prefix between indexes $a$ and $b$ in $\Psi$.

The suffix array and longest common prefix table can be constructed in $O\big((N_S + N_T) \log (N_S + N_T) \big)$ \cite{manber1993suffix}. Faster algorithms exist, but are not needed for our algorithm since it wouldn't affect our overall complexity. Also, note that $\operatorname{lcp}(a, b)$ can be computed in logarithmic time \cite{manber1993suffix}.

The precomputed sorted suffix array order is used to populate a persistent binary search tree $\tau$. The tree $\tau$ contains subsets of the indexes of $\Psi$ in natural order. We iterate through the suffixes of $S$ from largest to smallest. For each suffix $S_x$ we add $\Psi(S, x)$ to $\tau$. Since $\tau$ is a persistent tree, we can access any version of it \cite{sarnak1986planar}. This is a surprising property that is key to our algorithm. Further, it is worth noting that, due to persistence, storing all versions of the tree requires only $O(N_S \log N_S)$ space. Let $\tau_x$ be the version before inserting the suffix $x$. We can answer queries of the form ``what is the first element strictly before $z$'' and ``what is the first element strictly after $z$'' for a given index $z$. This can be computed by tree traversal in $O(\log N_S)$.

Now we show that the $\tau$ queries are sufficient to solve relaxed queries efficiently. Recall that a relaxed query $f(i,j,k)$ asks if there is a match of length $k$ starting at $i$ in $T$ and before $j$ in $S$. Consider $\Psi(T, i)$. The index of the longest match that starts in $S_{0}^{j-1}$ must be contained in $\tau_j$. Find the first entry $x$ less than $\Psi(T, i)$ and the first entry $y$ greater than $\Psi(T, i)$ in $\tau_j$. Since the entries of $\tau_j$ are sorted suffix array indexes, one of these two must be the largest match. This is because for a given sorted suffix array index $p$ in $\Psi$ the longest common prefix $\operatorname{lcp}(p, p+q)$ monotonically decreases as $q$ diverges from $0$. So, we can compute $\max(\operatorname{lcp}(\Psi(T, i), x), \operatorname{lcp}(\Psi(T, i), y))$ to answer our strict query. This costs $O(\log (N_S + N_T))$ time to search $\tau$ twice and then answer a longest common prefix query twice.

The overall complexity for a strict query is thus $O(\log (N_S + N_T) \log(\min(N_S, N_T)))$, since $R=O(\log (N_S + N_T))$. Our precomputation time is $O((N_S + N_T) \log (N_S + N_T))$ since it is dominated by suffix array construction. Define the number of $i,j$-pair queries as $Q$. The total complexity is  is dominated by the cost of answering all queries $O(Q \log (N_S + N_T) \log(\min(N_S, N_T)))$ when $Q$ is sufficiently large (e.g., $Q \geq N_T, N_S$). If we define $N=N_S+N_T$ a simple upper bound on the complexity of our algorithm is $O(N \log^2 N)$. This applies to many cases such as in the standard case where all queries have $i=j$.

\section{Experimental Results}

\begin{figure}[th!]
 \centering 
 \includegraphics[width = 0.9\columnwidth]{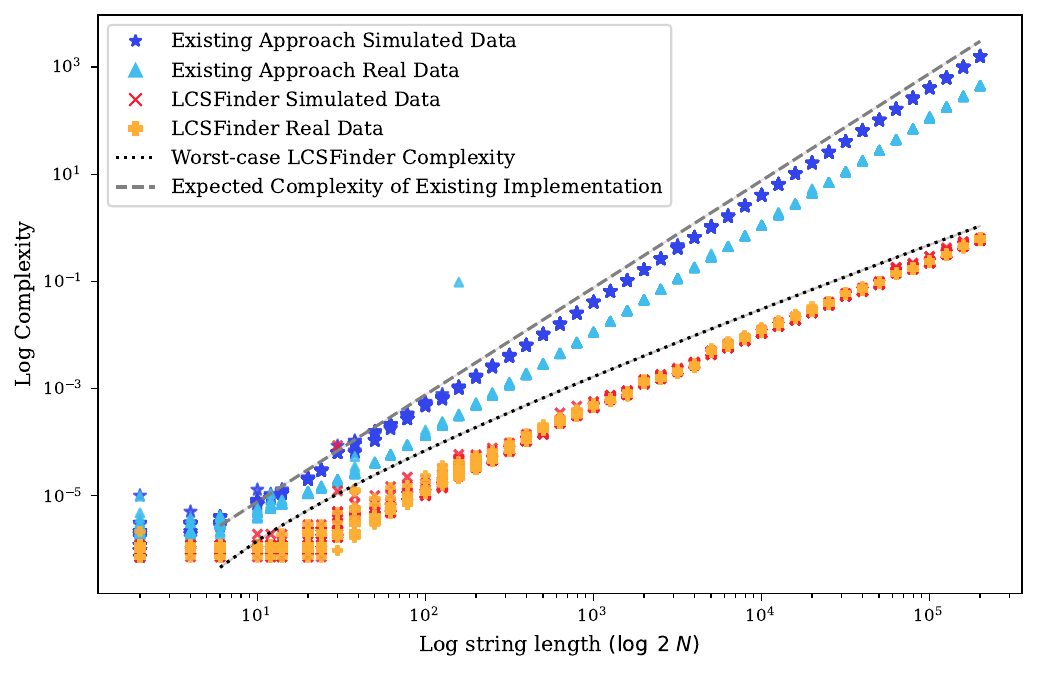}
 \caption{Empirically estimated complexity of the previously existing algorithm and that of LCSFinder as compared to the total length of both sequences ($N_T = N_S = N$). Note the log-log axes. The existing approach scales poorly as the sequence length increases due to its polynomial complexity. We can see that the LCSFinder algorithm achieves a performance bounded by the expected theoretical complexity. Similar performance is seen on the simulated and Twitter dataset, but LCSFinder's performance is closer for both.}
 \label{resultssuperfig}
\end{figure}

Worst-case analysis is helpful, but we are more interested in the typical performance, which governs the computation time in real applications. Expected performance is more difficult to find analytically because it depends on the structure of the data, and we are using non-parametric estimators in part because that structure is not known {\em a priori}. We validate the expected performance through empirical analysis on a simulated dataset (where we have control over the sequences) and a dataset containing a sample of real posts from the social media platform X (aka Twitter). 


\subsection{Validation on real and simulated datasets}
We evaluate on (i) synthetic sequences drawn i.i.d.\ from a discrete uniform distribution at varying lengths $N=10^i$ for $i=0.1,0.2,...,5.0$, and (ii) a preprocessed Twitter dataset of 2,846,284 tweets from 123 accounts \cite{southInformationFlowEstimation2022}. For each $N$ we truncate tokenised sequences and average 30 runs.

For each string length, the computation was repeated 30 times for each algorithm, with the results for each iteration shown in \autoref{resultssuperfig}. As the run times were measured using the wall-clock time, these times may have been affected by other processes performed by the computer. To mitigate this issue, a dedicated CPU was used.





\autoref{resultssuperfig} shows the results for simulations and real data. In the trials with both simulated and real data, we can see that the LCSFinder algorithm outperforms the existing approach across all sequence lengths, with the difference becoming more pronounced for longer sequences.

Note that for the most realistic sequence lengths (100,000 elements), LCSFinder offers an approximately 4 order of magnitude improvement. 
Moreover, the scaling in improvement means we can apply LCSFinder to desirable datasets much longer than the existing approach could handle. 

For short sequence lengths, the difference in run-time is smaller, but when entropy calculations are performed over many pairs of sequences (\eg when we are comparing many social media accounts) these computational advantages will still produce meaningful improvements to run-time. 

While the performance of the LCSFinder algorithm on real and simulated data is comparable, there is a larger difference for the existing implementation. This may be due to the shorter average match lengths in independent and uniformly sampled sequences, meaning fewer comparisons are required.  

The LCSFinder algorithm performs consistently between the real and simulated data, indicating that the expected complexity is achieved in many cases. Lines are used in \autoref{resultssuperfig} to indicate the expected complexity of the existing and LCSFinder algorithms, with the actual performance closely matching the expected performance for strings longer than 100 symbols for the proposed algorithm and sequences over 1000 for the existing approach. 


\section{Conclusion}

We propose the LCSFinder algorithm which provides the first sub-quadratic solution to the longest common substring matching problem when applied to entropy estimation. In the case of equal string lengths, we are able to achieve a worst-case complexity of $O(N \log^2 N)$. We speculate that replacing the suffix array data structure with a suffix tree may achieve a complexity of $O(N \log N)$. The suffix tree is a more powerful generalisation of a suffix array, and it may be the case that the binary search operation could be replaced by a tree traversal, speeding up the algorithm.

We validate the proposed LCSFinder algorithm on real and simulated data for strings up to $10^5$ symbols long. Results show that LCSFinder consistently outperforms existing methods across all sequence lengths, with behaviour matching expected complexity on both datasets. The proposed LCSFinder algorithm is more consistent across both real and simulated datasets. 

The LCSFinder algorithm is available as a standalone package and codebase \footnote{Package available at \url{https://pypi.org/project/LCSFinder}, codebase available at \url{https://github.com/bridget-smart/LCSFinder}}. It also includes a heuristic variant offering excellent practical performance with slower worst-case complexity, and has been integrated into the \pkg{ProcessEntropy} package.

 This contribution will allow non-parametric entropy estimation to be considered in cases where computational constraints were previously prohibitive, for example, on extremely large datasets, long sequences or for real-time calculations.


\bibliographystyle{IEEEtran}
\bibliography{Entropy_speed}
\end{document}